\documentclass[14pt,a4paper]{article}

\usepackage[cp1251]{inputenc}
\usepackage[english]{babel}
\usepackage{amsmath,amssymb,amsfonts,amsthm}
\usepackage{hyperref}
\usepackage{cite}
\usepackage[titletoc]{appendix}

\usepackage{times}
\usepackage{color}

\usepackage{indentfirst}

\begin{document}

\title{\textbf{Stable interactions between the extended Chern-Simons theory and a charged scalar field with higher derivatives: Hamiltonian formalism}}
\author{V.~A.~Abakumova\footnote{abakumova@phys.tsu.ru}, \ D.~S.~Kaparulin\footnote{dsc@phys.tsu.ru}, \ and S.~L.~Lyakhovich\footnote{sll@phys.tsu.ru}}
\date{\footnotesize\textit{Physics Faculty, Tomsk State University, Tomsk 634050, Russia}}

\maketitle

\begin{abstract}
\noindent We consider constrained multi-Hamiltonian formulation for the extended Chern-Simons theory with higher derivatives of arbitrary finite order. The order $n$ extension of the theory admits $(n-1)$-parametric series of conserved tensors. The $00$-component of any representative of the series can be chosen as Hamiltonian. The theory admits a series of Hamiltonian formulations, including the canonical Ostrogradski formulation. The Hamiltonian formulations with different Hamiltonians are not connected by canonical transformations. Also, we demonstrate the inclusion of stable interactions with charged scalar field that preserves one specified Hamiltonian from the series.
\end{abstract}

\section{Introduction}

The Hamiltonian formulations for theories with higher derivatives have been discussed once and again for decades since the work of Ostrogradski \cite{1}. The procedure for constructing such Hamiltonian formulation for degenerate theories was originally proposed in \cite{2}. This procedure and its modifications can be applied to study different gauge theories, including gravity (see \cite{3,4}). The canonical Ostrogradski Hamiltonian used in all these methods is not bounded. It leads to the well-known stability problem and difficulties with constructing of  quantum theory \cite{5,6,7}. 

The alternative Hamiltonian formulation was first introduced in the Pais-Uhlenbeck theory \cite{8,9}. In the work \cite{10}, it was noticed that a wide class of theories with higher derivatives admits the series of Hamiltonians and the Poisson brackets that are not connected with each other by canonical transformations. The canonical Ostrogradski Hamiltonian is included in the series. The series of non-canonical Hamiltonian formulations was constructed explicitly in \cite{11,12} for the extended Chern-Simons theory \cite{13} of the third and fourth order.

In the present work, we consider the Hamiltonian formulation for the extended Chern-Simons of arbitrary finite order. At free level, the theory of order $n$ admits $(n-1)$-parametric series of Hamiltonian formulations with Hamiltonian being the $00$-component of any conserved tensor from the series \cite{14}.  We demonstrate that non-Lagrangian interaction vertices with charged scalar field constructed in \cite{15} preserve one representative in the series of Hamiltonian formulations, and the theory remains stable at the interacting level.

The rest of the article is organized as follows. In section \ref{1}, we provide a general information about the extended Chern-Simons, including the equations of motion and conserved tensors. In section \ref{2}, we construct the multi-Hamiltonian formulation for the theory. The calculations here are mostly based on the techniques of working with Hankel and Bezout matrices that can be found in \cite{16}. In section \ref{3}, we consider the Hamiltonian formulation in the case of inclusion of interaction with charged scalar field.

\section{The extended Chern-Simons of order $n$}\label{1}

The extended Chern-Simons theory of order $n$ is a class of models of vector field $A=A_\mu(x)dx^\mu\,,\,\mu= 0\,, 1\,, 2\,,$ on $3d$ Minkowski space with the action functional
\begin{equation}\label{SChSnH} 
\displaystyle S[A(x)] = \frac{1}{2}\int{\ast A\wedge\Big(\sum\limits_{k=1}^{n}\alpha_k m^{2-k}(\ast d)^k A\Big)}\,,\qquad\ast dA = \varepsilon_{\mu\nu\rho}\partial^\mu A^\nu dx^\rho\,,\qquad\varepsilon_{012} = 1\,,
\end{equation}
where $m$ is a constant with dimension of mass, $\alpha_1\,,\,\ldots\,, \alpha_n$ are real numbers being the parameters of model, $\alpha_n\neq 0$\,, and $\ast\,, d$ denotes Hodge operator and de Rahm differential, respectively. The Minkowski metric is supposed to be mostly negative. The Lagrange equations for the action functional (\ref{SChSnH}) read
\begin{equation}\label{EChSnH} 
\displaystyle \frac{\delta S}{\delta A} = \Big(\sum\limits_{k=1}^{n} \alpha_k m^{2-k} (\ast d)^k\Big)A = 0\,.
\end{equation}
The action functional (\ref{SChSnH}) and equations of motion (\ref{EChSnH}) are invariant under the standard gradient gauge transformation for the field $A$\,.
 
The series of second-rank conserved tensors for the theory (\ref{SChSnH}) was constructed in \cite{14}. The most general representative of this series can be presented in the form
\begin{equation}\label{TmununH} 
\displaystyle T_{\mu\nu}(\alpha\,,\beta) = \frac{m^2}{2}\sum\limits_{r,s=1}^{n-1} C_{r,s}(\alpha\,,\beta)(F^{(r)}{}_\mu F^{(s)}{}_\nu + F^{(r)}{}_\nu F^{(s)}{}_\mu - \eta_{\mu\nu} \eta^{\rho\sigma}F^{(r)}{}_\rho F^{(s)}{}_\sigma)\,,
\end{equation}
where $\alpha=(\alpha_1\,,\,\ldots\,,\alpha_n)$ are parameters of the Lagrangian, and real numbers $\beta=(\beta_1\,,\,\ldots\,,\beta_n)$ are parameters of the series. We use the notation 
\begin{equation}\label{FrnH} 
\displaystyle F^{(r)} = \big(m^{-1}\ast d\big){}^r A\,,\qquad r = 1\,,\,\ldots\,, n-1\,,
\end{equation}
and $F^{(0)}{}_\mu\equiv A_\mu$\,. The square matrix $C_{r,s} (\alpha\,, \beta)\,,\, r\,, s = 1\,,\,\ldots\,,n\,,$ is determined by the generating relation
\begin{equation}\label{CrsH} 
\displaystyle \sum\limits_{r,s=1}^{n-1} C_{r,s}(\alpha\,,\beta)z^{r-1}u^{s-1} = \frac{M(z)N(u) - M(u)N(z)}{z-u}\,,
\end{equation}
where the polynomials of one variable $M(z), N(z)$ of the order $n-1$ read
\begin{equation}\label{MNnH} 
\displaystyle M(z) = \sum\limits_{r=0}^{n-1}\alpha_{r+1}z^r\,,\qquad N(z) = \sum\limits_{r=0}^{n-1}\beta_{r+1}z^r\,.
\end{equation}
The $C_{r,s}(\alpha\,, \beta)$ (\ref{CrsH}) is known as the Bezout matrix of the polynomials $M(z)\,, N(u)$ \cite{16}. Let us note that polynomial
 \begin{equation}\label{McharnH} 
\displaystyle M'(z) \equiv zM(z) = \sum\limits_{r=1}^{n}\alpha_rz^r\,,
\end{equation}
is so called characteristic polynomial of the theory (\ref{SChSnH}) \cite{14}. We get it by formal substitution of variable $z$ to the Lagrange equations (\ref{EChSnH}) instead of the Chern-Simons operator $\ast d$\,.

The conserved tensors of the model (\ref{SChSnH}) are defined as coefficients for parameters $\beta_1\,,\,\ldots\,,\beta_n$  in the series (\ref{TmununH}):
\begin{equation}\label{TmunurH} 
\displaystyle T^{(r)}{}_{\mu\nu} (\alpha) = \frac{\partial T_{\mu\nu}(\alpha\,,\beta)}{\partial\beta_r},\qquad r = 1,\ldots,n\,.
\end{equation}
By construction, $T^{(1)}{}_{\mu \nu}(\alpha)$ is the canonical energy-momentum for (\ref{SChSnH}), while $T^{(r)}{}_{\mu \nu}(\alpha)\,,\,r =  2\,,\,\ldots\,, n-1\,,$ are another independent conserved quantities. The quantity $T^{(n)}{}_{\mu \nu}(\alpha)$ (\ref{TmunurH}) is a linear combination of other conserved tensors by identity
\begin{equation}\label{} 
\displaystyle \sum\limits_{r=1}^{n}\alpha_r T^{(r)}{}_{\mu\nu}(\alpha) = 0\,.
\end{equation}
However, we consider $T^{(n)}{}_{\mu \nu}(\alpha)$ as it is convenient for the inclusion of interactions. Decomposition of an arbitrary representative of the series (\ref{TmununH}) in the basis independent generators $T^{(r)}{}_{\mu \nu}(\alpha)\,,\, r = 1\,,\,\ldots\,, n-1\,,$ (\ref{TmunurH}) reads
\begin{equation}\label{} 
\displaystyle T_{\mu\nu}(\alpha\,,\beta) = \frac{1}{\alpha_n}\sum\limits_{r=1}^{n-1}(\beta_r\alpha_n - \beta_n\alpha_r) T^{(r)}{}_{\mu\nu}(\alpha)\,.
\end{equation}
The canonical energy-momentum is always included in the series (\ref{TmununH}). It corresponds to the following values of the parameters:
\begin{equation}\label{betanH} 
\displaystyle  \beta_1 = 1\,,\qquad\beta_2 = \beta_3 = \ldots =\beta_n = 0\,,
\end{equation}
while other values of $\beta_1\,,\,\ldots\,, \beta_n$ in (\ref{TmununH}) define non-canonical conserved quantities.

The $00$-component $T_{00}(\alpha\,, \beta)$ of the conserved tensor (\ref{TmununH}) reads \begin{equation}\label{T00nH} 
\displaystyle T_{00}(\alpha\,,\beta) = \sum\limits_{r,s=1}^{n-1} C_{r,s}(\alpha\,,\beta)(F^{(r)}{}_i F^{(s)}{}_i + F^{(r)}{}_0 F^{(s)}{}_0)\,,
\end{equation}
where $i = 1\,, 2$\,, and the summation over repeated at the same level indices is implied. This quantity is a quadratic form of the variables (\ref{FrnH}), and it is positive definite if the matrix $C_{r,s}(\alpha, \beta)$ (\ref{CrsH}) is positive definite. A possible existence of bounded representatives in the series (\ref{T00nH}) is defined by the structure of roots of the characteristic polynomial: the bounded tensor exists if all the non-zero roots of the characteristic polynomial are real and different, and zero root has the multiplicity 1 or 2 \cite{14}. In terms of the polynomial $M(z)$ (\ref{MNnH}) it is sufficient to require that all its roots are real and different. The canonical energy of the theory (\ref{SChSnH}) is included in the series (\ref{T00nH}) with the parameters (\ref{betanH}), and it is always unbounded when $n>2$\,.

\section{Hamiltonian formulation of the extended Chern-Simons}\label{2}

In this section we show that the theory (\ref{EChSnH}) admits $(n-1)$-parametric series of the canonically inequivalent Hamiltonian formulations, and almost any representative in the series of conserved quantities (\ref{T00nH}) can be chosen as Hamiltonian. For doing this, we first lower the order of the equations (\ref{EChSnH}) to the first in time $t = x^0$ derivatives and then present the Poisson brackets and Hamiltonian that reduce this equations to the Hamiltonian form.

Let us introduce new variables that absorb time derivatives of the original vector field $A_\mu$\,, using spatial components of one-form $F^{(r)}{}_i\,, i = 1\,, 2\,,  \,r = 1\,,\,\ldots\,, n-1$ (\ref{FrnH}).  Then the first-order formulation for (\ref{SChSnH}) reads
\begin{equation}\label{F0iH} 
\displaystyle \partial_0 F^{(0)}{}_i = \partial_i A_0 - m\varepsilon_{ij} F^{(1)}{}_j\,;
\end{equation}
\begin{equation}\label{FriH} 
\displaystyle \partial_0 F^{(r)}{}_i = \frac{1}{m}\varepsilon_{ij}\partial_k\big(\partial_k F^{(r-1)}{}_j - \partial_j F^{(r-1)}{}_k\big) - m\varepsilon_{ij} F^{(r+1)}{}_j\,,\qquad r = 2\,,\,\ldots\,, n-2\,;
\end{equation}
\begin{equation}\label{Fn-1iH} 
\displaystyle \partial_0 F^{(n-1)}{}_i = \frac{1}{m}\varepsilon_{ij}\partial_k\big(\partial_k F^{(n-2)}{}_j - \partial_j F^{(n-2)}{}_k\big) + \frac{1}{\alpha_n}m\varepsilon_{ij}\sum\limits_{r=1}^{n-1}\alpha_r F^{(r)}{}_j\,;
\end{equation}
\begin{equation}\label{ThetaH} 
\displaystyle \mathit{\Theta} \equiv m\sum\limits_{r=1}^{n}\alpha_k\varepsilon_{ij}\partial_i F^{(r-1)}{}_j = 0\,.
\end{equation}
We can verify the equivalence of the equations (\ref{EChSnH}) and (\ref{F0iH}) -- (\ref{ThetaH}) as follows. The formulas (\ref{F0iH}), (\ref{FriH}) express the auxiliary  variables $F^{(1)}{}_i\,,\, i = 1\,, 2\,,\, r = 1\,,\,\ldots\,,n-1$\,, in terms of derivatives of $A$\,. After excluding all the variables, equations (\ref{Fn-1iH}) and (\ref{ThetaH}) reproduce spatial and temporal part of the equations of motion (\ref{EChSnH}), respectively. Let us note that in the first-order formalism the quantity $\mathit{\Theta}$ (\ref{ThetaH}) does not involve time derivatives and can be considered as a constraint.

The first-order system of equations (\ref{F0iH}) -- (\ref{Fn-1iH}) is Hamiltonian if there exists the Hamiltonian $H(\alpha\,, \beta)$ and Poisson brackets $\{\,,\,\}_{\alpha,\beta}$ such that
\begin{equation}\label{EqnH} 
\displaystyle \partial_0 F^{(r)}{}_i \approx \big\{\,F^{(r)}{}_i\,,\int dx\, H(\alpha\,,\beta)\,\big\}{}_{\alpha,\beta}\,,\qquad r = 0\,,\,\ldots\,, n-1\,.
\end{equation}
Here, the sign $\approx$ means that both sides of the equation are equal modulo the constraint $\mathit{\Theta}$ (\ref{ThetaH}). Introducing the parameters $\beta$\,, we take into account that different Hamiltonian formulations can exist for one and the same equations of motion. We choose the following ansatz for the Hamiltonian:
\begin{equation}\label{HnH} 
\displaystyle H(\alpha\,,\beta) = T_{00}(\alpha\,,\beta) + \Big(k_0A_0 + \frac{1}{m}\sum\limits_{r=1}^{n-1}k_r\varepsilon_{ij}\partial_i F^{(r-1)}{}_j\Big)\mathit{\Theta}\,.
\end{equation}
Here, $T_{00}(\alpha\,, \beta)$ is the $00$-component (\ref{T00nH}) of the conserved tensor (\ref{TmununH}) in terms of the phase-space variables:
\begin{equation}\label{} 
\displaystyle T_{00}(\alpha\,,\beta) = \frac{1}{2}m^2\sum\limits_{r,s=1}^{n-1} C_{r,s}(\alpha,\beta)\big(F^{(r)}{}_iF^{(s)}{}_i + \partial_iF^{(r-1)}{}_j(\partial_iF^{(s-1)}{}_j - \partial_jF^{(s-1)}{}_i)\big)\,,
\end{equation}
and $\mathit{\Theta}$ is the constraint (\ref{ThetaH}). We use the constants $k_0\,,k_1\,,\,\ldots\,,k_{n-1}$ in the Hamiltonian (\ref{HnH}) for the reasons of convenience and define them later. 

The Poisson bracket is determined from the condition that the equations (\ref{F0iH}) -- (\ref{Fn-1iH}) take the form (\ref{EqnH}) with the Hamiltonian (\ref{HnH}). It is equal to the following system of equations for the unknown Poisson bracket and parameters $k_0\,,k_1\,,\,\ldots\,,k_{n-1}$\,:
\begin{equation}\label{F0iHH} 
\displaystyle \big\{\,F^{(0)}{}_i\,,H(\alpha,\beta)\,\big\}{}_{\alpha,\beta} = \partial_iA_0 - m\varepsilon_{ij}F^{(1)}{}_j\,;
\end{equation}
\begin{equation}\label{FriHH} 
\displaystyle \big\{\,F^{(r)}{}_i\,,H(\alpha,\beta)\,\big\}{}_{\alpha,\beta} = \frac{1}{m}\varepsilon_{ij}\partial_k\big(\partial_kF^{(r-1)}{}_j - \partial_jF^{(r-1)}{}_k\big) - m\varepsilon_{ij}F^{(r+1)}{}_j\,,\qquad  r= 2\,,\,\ldots\,,n - 2\,;
\end{equation}
\begin{equation}\label{Fn-1iHH} 
\displaystyle \big\{\,F^{(n-1)}{}_i\,,H(\alpha,\beta)\,\big\}{}_{\alpha,\beta} = \frac{1}{m}\varepsilon_{ij}\partial_k\big(\partial_kF^{(n-2)}{}_j - \partial_jF^{(n-2)}{}_k\big) + \frac{1}{\alpha_n}m\varepsilon_{ij}\sum\limits_{r=1}^{n-1}\alpha_rF^{(r)}{}_j\,.
\end{equation}
These relations are linear equations for the unknown elements of the Poisson brackets matrix of the phase-space variables $F^{(1)}{}_i\,,\,i = 1\,, 2\,,\, r = 0\,,\,\ldots\,,n-1$\,. In a class of field independent Poincar\'e-invariant Poisson brackets the system has the following solution:\begin{equation}\label{Fn-1ijH} 
\displaystyle \big\{\,F^{(n-1)}{}_i(\vec{x})\,,F^{(n-1)}{}_j(\vec{y})\,\big\}{}_{\alpha,\beta} = \frac{1}{\alpha_nm\,\text{det}\,C(\alpha\,,\beta)}\Big(\sum\limits_{r=1}^{n-1}\alpha_rM^{r,n-1}(\alpha\,,\beta)\Big)\varepsilon_{ij}\delta(\vec{x}-\vec{y})\,;
\end{equation}
\begin{equation}\label{FrijH} 
\displaystyle \big\{\,F^{(r)}{}_i(\vec{x})\,,F^{(s)}{}_j(\vec{y})\,\big\}{}_{\alpha,\beta} = - \frac{M^{r,s+1}(\alpha\,,\beta)}{m\,\text{det}\,C(\alpha\,,\beta)}\varepsilon_{ij}\delta(\vec{x}-\vec{y})\,,\qquad r\,,s+1 = 1\,,\,\ldots\,,n-1\,;
\end{equation}
\begin{equation}\label{AijH} 
\displaystyle \big\{\,A_i(\vec{x})\,,A_j(\vec{y})\,\big\}{}_{\alpha,\beta} = - \frac{\gamma}{m\,\text{det}\,C(\alpha\,,\beta)}\varepsilon_{ij}\delta(\vec{x}-\vec{y})\,.
\end{equation}
The parameters $k_0\,,\,k_1\,,\ldots\,,\,k_{n-1}$ are defined by the formulas
\begin{equation}\label{}
\begin{array}{c}
\displaystyle k_0 = - \frac{\text{det}\,C(\alpha\,,\beta)}{\alpha_1\gamma + \sum\limits_{s=2}^{n}\alpha_sM^{s-1,1}(\alpha\,,\beta)}\,,\\[3mm]
\displaystyle k_r = \frac{\gamma C_{r,1}(\alpha\,,\beta) + \sum\limits_{s=2}^{n-1}C_{r,s}(\alpha\,,\beta)M^{s-1,1}(\alpha\,,\beta)}{\alpha_1\gamma + \sum\limits_{s=2}^{n}\alpha_sM^{s-1,1}(\alpha\,,\beta)}\,,\qquad r = 1\,,\,\ldots\,,n-1\,.
\end{array}
\end{equation}
In (\ref{Fn-1ijH}) -- (\ref{AijH}), the quantity $\gamma$ is a free parameter, and $M^{r,s}{}(\alpha\,,\beta)$ denotes adjugate for the Bezout matrix $C_{r,s}{}(\alpha\,,\beta)$ (\ref{CrsH}):
\begin{equation}\label{} 
\displaystyle \sum\limits_{k=1}^{n-1}C_{r,k}(\alpha\,,\beta)M^{k,s}(\alpha\,,\beta) = \text{det}\,C(\alpha\,,\beta)\delta_r{}^s\,.
\end{equation}
The solution (\ref{Fn-1ijH}) -- (\ref{AijH}) of the equations (\ref{F0iHH}) -- (\ref{Fn-1iHH}) is well-defined, if
\begin{equation}\label{condneq0H} 
\displaystyle \text{det}\,C(\alpha\,,\beta)\neq0\,,\qquad \alpha_1\gamma + \sum\limits_{s=2}^{n}\alpha_sM^{s-1,1}(\alpha\,,\beta)\neq0\,.
\end{equation}
The compatibility conditions (\ref{condneq0H}) for the equations (\ref{F0iHH}) -- (\ref{Fn-1iHH}) have a simple physical interpretation. The first relation is equivalent to the nondegeneracy of the Hamiltonian quadratic form (\ref{HnH}). The second relation guarantees that the constraint $\mathit{\Theta}$ (\ref{ThetaH}) generates gauge symmetries for the vector potential $A$\,.

We use the relations 
\begin{equation}\label{} 
\displaystyle M^{r,s}(\alpha\,,\beta) - M^{k,l}(\alpha\,,\beta) = 0\,,\qquad s + r = k + l\,;
\end{equation}
\begin{equation}\label{} 
\displaystyle \sum\limits_{k=1}^{n-1}\alpha_kM^{r,k}(\alpha\,,\beta) + \alpha_nM^{r+1,n}(\alpha\,,\beta) = 0\,,\qquad r = 1\,,\,\ldots\,,n-2\,,
\end{equation}
to verify that formulas (\ref{Fn-1ijH}) -- (\ref{AijH}) represent the solution of the equations (\ref{F0iHH}) -- (\ref{Fn-1iHH}). These conditions are satisfied, because adjugate for the Bezout matrix $C_{r,s}{}(\alpha\,, \beta)$ (\ref{CrsH}) is the Hankel matrix constructed from polynomials (\ref{MNnH}). The proofs of this relations can be found in \cite{16}. 

Formulas (\ref{EqnH}), (\ref{HnH}), (\ref{Fn-1ijH}) -- (\ref{AijH}) define a series of Hamiltonian formulations for the extended Chern-Simons theory (\ref{SChSnH}). An arbitrary representative in this series is determined by $2n+1$ parameters $\alpha_1\,,\,\ldots\,,\alpha_n\,,\,\beta_1\,,\,\ldots\,,\beta_n\,,\,\gamma$\,. The constants $\alpha_1\,,\,\ldots\,,\alpha_n$ are parameters of the model (\ref{SChSnH}), and real numbers $\beta_1\,,\,\ldots\,,\,\beta_{n-1}$ define a representative in the series of the conserved quantities (\ref{TmununH}) that will be chosen as Hamiltonian. The quantities $\beta_n\,,\,\gamma$ are auxiliary parameters: $\beta_n$ can be always absorbed by redefining parameters $\beta_1\,,\,\ldots\,,\beta_{n-1}$\,, and constant $\gamma$ determine a specific representative in the equivalence class of the Poisson brackets (\ref{Fn-1ijH}) -- (\ref{AijH}). The Poisson brackets for the physical observables do not depend on the value of $\gamma$\,. The total number of the independent parameters that lead to the inequivalent Hamiltonian formulation of the model (\ref{SChSnH}) is $n-1$\,, so the extended Chern-Simons theory admits $(n-1)$-parametric series of Hamiltonian formulations.

For all the admissible parameters in the Hamiltonian, the Poisson bracket is a nondegenerate tensor
\begin{equation}\label{} 
\displaystyle \text{det}\,\big\{\,F^{(r)}{}_i(\vec{x})\,,F^{(s)}{}_j(\vec{y})\,\big\}{}_{\alpha,\beta} = - \frac{1}{\alpha_n\,\text{det}^2\,C(\alpha\,,\beta)}\big(\alpha_1\gamma + \sum\limits_{r=2}^{n}\alpha_rM^{r-1,1}(\alpha\,,\beta)\big)\neq 0\,.
\end{equation}
In this case, the Hamilton's equations (\ref{F0iH}) -- (\ref{ThetaH}) follow from the variational principle for the functional 
\begin{equation}\label{SH} 
\displaystyle S(\alpha\,,\beta) = \int\,d^3x\,\Big(m\sum\limits_{r,s=0}^{n-1}\Omega_{r,s}(\alpha\,,\beta)\varepsilon_{ij}F^{(r)}{}_i\partial_0F^{(s)}{}_i - H(\alpha\,,\beta)\Big)\,.
\end{equation}
The symplectic form $\Omega_{r\,,s}{}(\alpha\,,\beta)$ is defined by the generating relation
\begin{equation}\label{OmegaH} 
\displaystyle \sum\limits_{r,s=0}^{n-1}\Omega_{r,s}(\alpha\,,\beta)z^ru^s = - \frac{\text{det}\,C(\alpha\,,\beta)}{\alpha_1\gamma + \sum\limits_{s=2}^{n}\alpha_sM^{s-1,1}(\alpha\,,\beta)} \frac{M'(z)N'(u) - M'(u)N'(z)}{z-u}\,,
\end{equation}
where $M'(z)$ is the characteristic polynomial (\ref{McharnH}) of the theory (\ref{SChSnH}), and $N'(z)$ is given by the formula
\begin{equation}\label{NcharnH} 
\displaystyle N'(z) = \beta_1 + \sum\limits_{r=1}^{n-1}\Big(\beta_{r+1} - \frac{1}{\text{det}\,C\,(\alpha\,,\beta)}\Big(\beta_1\gamma + \sum_{k=2}^{n}\beta_kM^{k-1,1}(\alpha\,,\beta)\Big)C_{1,r}(\alpha\,,\beta)\Big)z^r\,.
\end{equation}
To arrive to the relations (\ref{OmegaH}), (\ref{NcharnH}), we need to use the inversion formula for the Hankel matrix of the Poisson brackets  from \cite{16}. The formulas (\ref{SH}) -- (\ref{NcharnH}) allow us to systematically reproduce the symplectic structure if the Hamiltonian of the theory is given.

The canonical Ostrogradski Hamiltonian formulation \cite{1} is reproduced by the formulas (\ref{HnH}), (\ref{SH}) -- (\ref{OmegaH}) with the following values of the parameters in the Hamiltonian:
\begin{equation}\label{} 
\displaystyle \beta_1 = 1\,,\qquad \beta_2 = \beta_3 =\ldots = \beta_n =0\,,\qquad\gamma = 0\,.
\end{equation}
In this case, the first-order action takes the form
\begin{equation}\label{ScanH} 
\displaystyle S(\alpha\,,\beta) = \int\,d^3x\Big(m\sum\limits_{r,s=0}^{n-1}\alpha_{s+r+1}\varepsilon_{ij}F^{(r)}{}_i\partial_0F^{(s)}{}_i - T^{(0)}{}_{00}(\alpha) - A_0\mathit{\Theta}\Big)\,,
\end{equation}
where $T^{(0)}{}_{00}(\alpha)$ is the $00$-component of the canonical energy-momentum, and it is assumed that $\alpha_r = 0$ for all $r>n$\,.  It is obvious, that canonical Hamiltonian action (\ref{ScanH}) is not equivalent to the general representative in the series (\ref{SH}), because the canonical Hamiltonian is always unbounded, while in general the bounded representatives are admissible.

\section{Couplings with charged scalar}\label{3}

In \cite{16}, we suggest the following interaction vertices for the extended Chern-Simons theory (\ref{SChSnH}) and charged scalar field $\varphi = \text{Re}\,\varphi(x) + i\,\text{Im}\,\varphi(x)$ with higher derivatives:
\begin{equation}\label{EqintH} 
\displaystyle \Big(\sum\limits_{k=1}^{n}\alpha_km^{2-k}(\ast d)^k\Big)A - \sum\limits_{a=1}^{N}ie_aj^{(a)}(\beta\,;\varphi\,,A) = 0\,,\qquad\Big(\prod\limits_{a=1}^{N}\big(D_\mu D^\mu + \rho^am^2\big)\Big)\varphi = 0\,.
\end{equation}
Here, $\beta_1\,,\,\ldots\,,\beta_n\,,\,e_1\,,\,\ldots\,,e_N$ are coupling constants, and we use the following notation:
\begin{equation}\label{japhiaH} 
\displaystyle j^{(a)}(\beta\,;\varphi\,,A) = i\big(\varphi^{(a)}(D_\mu \varphi^{(a)}){}^\ast - \varphi^{(a)}{}^\ast(D_\mu\varphi^{(a)})\big)\,,\qquad\varphi^{(a)} = \Big(\prod\limits_{b\neq a}\frac{D_\mu D^\mu + \rho^bm^2}{(\rho^a - \rho^b)m^2}\Big)\varphi\,.
\end{equation}
The parameters $\rho^a\,,\,a = 1\,,\,\ldots\,, N$\,, for the complex scalar field are non-negative and pairwise distinct. The covariant derivative is defined by the non-minimal way:
\begin{equation}\label{} 
\displaystyle D_\mu\varphi = \Big(\partial_\mu - i\,m\sum\limits_{r=1}^{n}\beta_rF^{(r-1)}{}_\mu\Big)\varphi\,.
\end{equation}
The covariant derivative for the complex conjugate scalar field is given by the complex conjugation of this relation. The gauge symmetry for the theory (\ref{EqintH}) reads
\begin{equation}\label{} 
\displaystyle \delta_ \xi A_\mu(x) = \partial_\mu\xi(x)\,,\qquad \delta_\xi\varphi(x) = - i\beta_1\varphi^\ast(x)\,.
\end{equation}
The conserved tensor for (\ref{EqintH}) is given by the relation
\begin{equation}\label{} 
\displaystyle \mathit{\Theta}_{\mu\nu}(\alpha\,,\beta) = T_{\mu\nu}(\alpha\,,\beta) + \sum_{a=1}^{N}e_a\big(D_\mu\varphi(D^\mu\varphi)^\ast + \rho^am^2\varphi\,\varphi^\ast\big),
\end{equation}
where $T_{\mu\nu}(\alpha\,,\beta)$ denotes the quantity (\ref{TmununH}).

The first-order formulation of the theory (\ref{EqintH}) reads
\begin{equation}\label{F0intH} 
\displaystyle \partial_0 F^{(0)}{}_i = \partial_i A_0 - m\varepsilon_{ij} F^{(1)}{}_j\,;
\end{equation}
\begin{equation}\label{FrintH} 
\displaystyle \partial_0 F^{(r)}{}_i = \frac{1}{m}\varepsilon_{ij}\partial_k\big(\partial_k F^{(r-1)}{}_j - \partial_j F^{(r-1)}{}_k\big) - m\varepsilon_{ij} F^{(r+1)}{}_j\,,\qquad r = 2\,,\ldots\,, n-2\,;
\end{equation}
\begin{equation}\label{Fn-1intH} 
\displaystyle \partial_0 F^{(n-1)}{}_{i} = \frac{1}{m}\varepsilon_{ij}\partial_k\big(\partial_k F^{n-2}{}_j - \partial_j F^{(n-2)}{}_k\big) + \frac{1}{\alpha_n}m\varepsilon_{ij}\sum\limits_{r=1}^{n-1}\alpha_r F^{(r)}{}_{j} - \frac{1}{\alpha_nm}\sum\limits_{a=1}^{N}ie_a(j_a)_i(\beta\,;\varphi\,,A)\,;
\end{equation}
\begin{equation}\label{phiaintH} 
\displaystyle \partial_0\varphi^{(a)} = \pi^{(a)}{}^\ast + i\Big(\beta_1 A_0 + \frac{1}{m}\sum\limits_{r=2}^{n}\beta_r\varepsilon_{ij}\partial_i F^{(r-2)}{}_{j}\Big)\varphi^{(a)}\,,\qquad a = 1\,,\,\ldots,N + c.c.\,;
\end{equation}
\begin{equation}\label{piaintH} 
\displaystyle \partial_0\pi^{(a)} = (D_iD_i -\rho_am^2)\varphi^{(a)} + i\Big(\beta_1 A_0 + \frac{1}{m}\sum\limits_{r=2}^{n}\beta_r\varepsilon_{ij}\partial_i F^{(r-2)}{}_j\Big)\pi^{(a)}\,,\qquad a = 1\,,\ldots\,,N + c.c.\,;
\end{equation}
\begin{equation}\label{ThetaintH} 
\displaystyle \mathit{\Theta}\equiv m\Big(\sum\limits_{r=1}^{n}\alpha_r\varepsilon_{ij}\partial_i F^{r-1}{}_{j} - \sum\limits_{a=1}^{N}ie_a\big(\varphi^{(a)}\pi^{(a)} - (\varphi^{(a)}){}^\ast(\pi^{(a)})^\ast\big)\Big) = 0\,.
\end{equation}
Here, $F^{(r)}{}_i\,,\,i = 1\,, 2\,,\, r = 1\,,\,\ldots\,,n-1$ (\ref{FrnH}),\,$\varphi^{(a)}\,,\,a =1\,,\,\ldots\,, N$ (\ref{japhiaH}), and $\pi^{(a)}\,,\,a = 1\,,\,\ldots\,,N$\,, are the new additional variables that absorb time derivatives of the original vector and scalar fields.  All the additional variables are excluded from the equations (\ref{F0intH}) -- (\ref{ThetaintH}), and after that the resulting system coincides with the original higher derivative equations. Let us note that in the first-order formalism $ \mathit{\Theta}$ (\ref{ThetaintH}) does not involve higher derivatives and can be considered as the constraint.

The equations (\ref{F0intH}) -- (\ref{piaintH}) are Hamiltonian in the sense of (\ref{EqnH}) with respect to the Hamiltonain
\begin{equation}\label{HintH} 
\begin{array}{c}
\displaystyle H(\alpha\,,\beta) = T_{00}(\alpha\,,\beta) + \sum\limits_{a=1}^{N}e_a(\pi^{(a)}(\pi^{(a)})^\ast + D_i\varphi(D_i\varphi)^\ast + \rho^am^2\varphi\,\varphi^\ast)\,+\\[3mm]
\displaystyle +\,(\beta_1 A_0 + \frac{1}{m}\sum\limits_{p=1}^{n-1}\beta_{p+1}\varepsilon_{ij}\partial_i F^{(p-1)}{}_j)\mathit{\Theta}\,,
\end{array}
\end{equation}
where $T_{00}(\alpha, \beta)$ is defined by (\ref{T00nH}). The Poisson bracket of the phase-space variables is determined by
\begin{equation}\label{Fn-1ijintH} 
\displaystyle \big\{\,F^{n-1}{}_i(\vec{x})\,, F^{(n-1)}_{j}(\vec{y})\,\big\}{}_{\alpha,\beta} = \frac{1}{\alpha_nm\,\text{det}\,C(\alpha\,,\beta)}\Big(\sum\limits^{n-1}_{r=1}\alpha_rM^{r-1,n-1}(\alpha\,,\beta)\Big)\varepsilon_{ij}\delta(\vec{x}-\vec{y})\,;
\end{equation}
\begin{equation}\label{FrijintH} 
\displaystyle \big\{\,F^{(r)}_i(\vec{x})\,, F^{(s)}_j(\vec{y})\,\big\}{}_{\alpha,\beta} = - \frac{M^{r,s-1}(\alpha\,,\beta)}{m\,\text{det}\,C(\alpha\,,\beta)}\varepsilon_{ij}\delta(\vec{x}-\vec{y})\,,\qquad r\,,\,s-1 = 1\,,\,\ldots\,,n-1\,;
\end{equation}
\begin{equation}\label{AijintH} 
\displaystyle \big\{\,A_i(\vec{x})\,,A_j(\vec{y})\,\big\}{}_{\alpha,\beta} = - \frac{1}{\beta_1m\,\text{det}\,C(\alpha\,,\beta)}\Big(\sum\limits_{k=1}^{n-1}\beta_{k+1}M^{l,k}(\alpha\,,\beta)\Big)\varepsilon_{ij}\delta(\vec{x}-\vec{y})\,;
\end{equation}
\begin{equation}\label{phipiintH} 
\displaystyle \big\{\,\varphi^{(a)}(\vec{x})\,,\pi^{(b)}(\vec{y})\,\big\}{}_{\alpha,\beta} = \frac{1}{e_a}\,\delta^{ab}\,\delta(\vec{x}-\vec{y})\,.
\end{equation}
The relations (\ref{Fn-1ijintH}) -- (\ref{phipiintH}) are well-defined, if
\begin{equation}\label{} 
\displaystyle \text{det}\,C(\alpha\,,\beta) \neq0\,,\qquad e_a\neq0\,,\qquad \beta_1\neq0\,.
\end{equation}
The first two of these conditions guarantee that the Hamiltonian is nondegenerate in a free limit, while the last one provides the existence of gauge $U(1)$-transformation for the scalar field. A case with $\beta_1 = 0$ it is supposed to be special and is not considered in this work. So, it is shown that almost all the constructed in \cite{15} interactions admit Hamiltonian formulation. This Hamiltonian formulation is canonically inequivalent to the Ostrogradski formulation, as all the original equations are non-Lagrangian.

Let us illustrate the general construction of the Hamiltonian formulation using the extended Chern-Simons of the third order coupling to the massless charged scalar. In this case, the equations of motion (\ref{EqintH}) read
\begin{equation}\label{Eqint3H} 
\displaystyle \Big(\frac{1}{m}\alpha_3(\ast d){}^3 + \alpha_2(\ast d){}^2 + m\alpha_1(\ast d)\Big)A + ie\big(\varphi^\ast(D_\mu\varphi) - \varphi(D_\mu\varphi){}^\ast\big)dx^\mu = 0\,,\qquad D_\mu D^\mu\varphi = 0\,,
\end{equation}
where $e = e_1$ is a coupling constant, and the covariant derivative is defined by the relation
\begin{equation}\label{} 
\displaystyle D_\mu\varphi = \big(\partial_\mu - i(\beta_1A_\mu + \beta_2F_\mu + \beta_3G_\mu)\big)\varphi\,,\qquad F_\mu\equiv\frac{1}{m}(\ast d A)_\mu\,,\qquad G_\mu\equiv\frac{1}{m^2}\big((\ast d){}^2A\big){}_\mu\,.
\end{equation}
The first-order formulation (\ref{F0intH}) -- (\ref{ThetaintH}) for equations (\ref{Eqint3H}) can be written in a form:
\begin{equation}\label{Ai3H} 
\displaystyle \partial_0A_i = \partial_iA_0 - m\varepsilon_{ij}F_j\,;
\end{equation}
\begin{equation}\label{Fi3H} 
\displaystyle \partial_0F_i = \frac{1}{m}\varepsilon_{ij}\partial_k\big(\partial_kA_j - \partial_jA_k\big) - m\varepsilon_{ij}G_j\,;
\end{equation}
\begin{equation}\label{Gi3H} 
\displaystyle \partial_0G_i = \frac{1}{m}\varepsilon_{ij}\partial_k\big(\partial_kF_j - \partial_jF_k\big) + \frac{1}{\alpha_n}\Big(m\varepsilon_{ij}\big(\alpha_1F_j + \alpha_2G_j\big) + \frac{1}{m}ie\big(\varphi^\ast(D_i\varphi) - \varphi(D_i\varphi){}^\ast\big)\Big)\,;
\end{equation}
\begin{equation}\label{phi3H} 
\displaystyle \partial_0\varphi = \pi^\ast + i\Big(\beta_1A_0 + \frac{1}{m}\big(\beta_2\varepsilon_{ij}\partial_iA_j + \beta_3\varepsilon_{ij}\partial_iF_j\big)\Big)\varphi\,,\qquad + c.c.\,;
\end{equation}
\begin{equation}\label{pi3H} 
\displaystyle \partial_0\pi = D_iD_i\varphi^\ast -i\Big(\beta_1A_0 + \frac{1}{m}\big(\beta_2\varepsilon_{ij}\partial_iA_j + \beta_3\varepsilon_{ij}\partial_iF_j\big)\Big)\pi\,,\qquad+ c.c.\,;
\end{equation}
\begin{equation}\label{Theta3H} 
\displaystyle \mathit{\Theta}\equiv m\varepsilon_{ij}\big(\alpha_1\partial_iA_j + \alpha_2\partial_iF_j +\alpha_3\partial_iF_j\big) + ie\big(\varphi\pi - \varphi^\ast\pi^\ast\big) = 0\,.
\end{equation}
Then the Hamiltonian (\ref{HintH}) reads
\begin{equation}\label{H3H} 
\begin{array}{c}
\displaystyle H(\alpha,\beta) = \frac{m^2}{2}\Big[(\beta_2\alpha_3 -\beta_3\alpha_2)\big(G_iG_i + \partial_iF_j(\partial_iF_j - \partial_jF_i)\big) + 2(\beta_1\alpha_3 - \beta_3\alpha_1)\big(G_iF_i\,+ \\[3mm]
\displaystyle +\,\partial_iF_j(\partial_iA_j - \partial_jA_i)\big) + (\beta_1\alpha_2 - \beta_2\alpha_1)\big(F_iF_i + \partial_iA_j(\partial_iA-j - \partial_jA_i)\big)\Big]\,+\\[3mm]
\displaystyle +\,e\big(\pi\pi^\ast + D_i\varphi(D_i\varphi){}^\ast\big) + \Big(\beta_1A_0 + \frac{1}{m}\big(\beta_2\varepsilon_{ij}\partial_iA_j + \beta_3\varepsilon_{ij}\partial_iF_j\big)\Big)\mathit{\Theta}\,.
\end{array}
\end{equation}
The Poisson brackets (\ref{Fn-1ijintH}) -- (\ref{phipiintH}) of the phase-space variables are given by the relations
\begin{equation}\label{Gij3H} 
\displaystyle \big\{\,G_i(\vec{x})\,,G_j(\vec{y})\,\big\}{}_{\alpha\,,\beta} = \frac{\beta_3\alpha^2_1 -\beta_2\alpha_2\alpha_1 + \beta_1(\alpha^2_2 - \alpha_3\alpha_1)}{m\,\text{det}\,C(\alpha\,,\beta)}\varepsilon_{ij}\delta(\vec{x}-\vec{y})\,;
\end{equation}
\begin{equation}\label{GFij3H} 
\displaystyle \big\{\,G_i(\vec{x})\,,F_j(\vec{y})\,\big\}{}_{\alpha,\beta} = \frac{\beta_1\alpha_2 -\beta_2\alpha_1}{m\,\text{det}\,C(\alpha\,,\beta)}\varepsilon_{ij}\delta(\vec{x}-\vec{y})\,;
\end{equation}
\begin{equation}\label{FFij3H} 
\displaystyle \big\{\,F_i(\vec{x})\,,F_j(\vec{y})\,\big\}{}_{\alpha,\beta} = \big\{\,G_i(\vec{x})\,,A_j(\vec{y})\,\big\}{}_{\alpha,\beta} = \frac{\beta_1\alpha_3 - \beta_3\alpha_1}{m\,\text{det}\,C(\alpha\,,\beta)}\varepsilon_{ij}\delta(\vec{x}-\vec{y})\,;
\end{equation}
\begin{equation}\label{FAij3H} 
\displaystyle \big\{\,F_i(\vec{x})\,,A_j(\vec{y})\,\}{}_{\alpha,\beta} = \frac{\beta_3\alpha_2 -\beta_2\alpha_3}{m\,\text{det}\,C(\alpha\,,\beta)}\varepsilon_{ij}\delta(\vec{x}-\vec{y})\,;
\end{equation}
\begin{equation}\label{Aij3H} 
\displaystyle \big\{\,A_i(\vec{x})\,,A_j(\vec{y})\,\big\}{}_{\alpha\,,\beta} = \frac{\beta^2_3\alpha_1 -\beta_3\beta_2\alpha_2 -\beta_3\beta_1\alpha_3 +\beta^2_2\alpha_3}{\beta_1m\,\text{det}\,C(\alpha\,,\beta)}\varepsilon_{ij}\delta(\vec{x}-\vec{y})\,;
\end{equation}
\begin{equation}\label{phipi3H} 
\displaystyle \big\{\,\varphi(\vec{x})\,,\pi(\vec{y})\,\big\}{}_{\alpha,\beta} = \frac{1}{e}\,\delta(\vec{x}-\vec{y})\,.
\end{equation}
Here, we use the notation
\begin{equation}\label{} 
\displaystyle \text{det}\,C(\alpha\,,\beta)\equiv\beta^2_3\alpha^2_1 - \beta_3\beta_2\alpha_2\alpha_1 + \beta_3\beta_1(2\alpha^2_2 -2\alpha_3\alpha_1) + \beta^2_2\alpha_3\alpha_1 - \beta_2\beta_1\alpha_3\alpha_1 + \beta^2_1\alpha^2_3\,.
\end{equation}
The Hamiltonian (\ref{H3H}) and Poisson brackets (\ref{Gij3H}) -- (\ref{phipi3H}) are well-defined, whenever
\begin{equation}\label{} 
\displaystyle \text{det}\,C(\alpha\,,\beta)\neq0\,,\qquad \beta_1\neq0\,,\qquad e\neq0\,.
\end{equation}
Then $\varphi = \pi = e = 0$\,, the vector field dynamics split off, and formulas (\ref{Ai3H}) -- (\ref{Gi3H}), (\ref{Gij3H}) -- (\ref{Aij3H}) reproduce one of the admissible representatives in the series of Hamiltonian formulations for free extended Chern-Simons theory of the third order \cite{11}\,. This establishes correspondence with the previous results.

\section{Conclusion}
We have seen that at free level the extended Chern-Simons theory of order $n$ is multi-Hamiltonian, and it admits $(n-1)$-parametric series of non-canonical Hamiltonian formulations with $00$-component of any representative of $(n-1)$-parametric series of conserved tensors as the Hamiltonian. For the certain range of the parameters of the theory, there exist the bounded representatives, otherwise all the Hamiltonians in the series are unbounded.
The stability condition is defined by the positive definiteness of the matrix $C_{r,s}(\alpha, \beta)$ (\ref{CrsH}). The series includes the canonical Hamiltonian formulation with always unbounded Ostrogradski Hamiltonian. Only one of these Hamiltonians survives then we include the non-Lagrangian interactions with charged scalar \cite{15}. Its parameters are defined by the values of the coupling constants. It allows us to preserve the stability and consistently quantize the theory at the interacting level.
This work was conducted within a government task of the Ministry of Education and Science of the Russian Federation, project No.3.9594.2017/8.9.


\begin{thebibliography}{11}
\bibitem{1} M.~V.~Ostrogradski, Memoires sur les equations differentielles relatives au probleme des isoperimetres, Mem. Acad. St. Petersburg. \textbf{6}, (1850) 385-517.
\bibitem{2} D.~M.~Gitman, S.~L.~Lyakhovich, I.~V.~Tyutin, Hamilton formulation of a theory with high derivatives, Sov. Phys. J. \textbf{26}, (1983) 61-66.
\bibitem{3} J.~Kluson, M.~Oksanen, A.~Tureanu, Hamiltonian analysis of curvature-squared gravity with or without conformal invariance, Phys. Rev. D \textbf{89}, (2014) 064043.
\bibitem{4} Y.~Ohkuwa, Y.~Ezawa, On the canonical formalism of f(R)-type gravity using Lie derivatives, Eur. Phys. J. Plus \textbf{77}, (2015) 130.
\bibitem{5} E.~T.~Tomboulis, Renormalization and unitary in higher derivative and nonlocal gravity theories, Mod. Phys. Lett. A \textbf{30}, (2015) 1540005.
\bibitem{6} M.~Pavsic, Pais-Uhlenbeck oscillator and negative energies, Int. J. Geom. Methods Mod. Phys. \textbf{13}, (2016) 1630015.
\bibitem{7} A.~V.~Smilga, Classical and quantum dynamics of higher-derivative systems, Int. J. Mod. Phys. A \textbf{32}, (2017) 1730025.
\bibitem{8} K.~Bolonek, P.~Kosinski, Hamiltonian structures for Pais-Uhlenbeck oscillator, Acta Phys. Polon. B \textbf{36}, (2005) 2115.
\bibitem{9} E.~V.~Damanskinsky, M.~A.~Sokolov, Remarks on quantization of Pais-Uhlenbeck oscillators, J. Phys. A: Math. Gen. \textbf{39}, (2006) 10499.
\bibitem{10} D.~S.~Kaparulin, S.~L.~Lyakhovich, A.~A.~Sharapov, Classical and quantum stability of higher-derivative dynamics, Eur. Phys. J. C \textbf{74}, (2017) 3072.
\bibitem{11} V.~A.~Abakumova, D.~S.~Kaparulin, S.~L.~Lyakhovich, Multi-Hamiltonian formulations and stability of higher-derivative extensions of 3d Chern-Simons, Eur. Phys. J. C \textbf{78}, (2018) 115.
\bibitem{12} V. ~A. ~Abakumova, D.~S.~Kaparulin, S.~L.~Lyakhovich, Bounded Hamiltonian in the fourth-order extension of the Chern-Simons theory, Russ. Phys. J. \textbf{60}, (2018) 2095-2104.
\bibitem{13} S.~Deser, R.~Jackiw, Higher derivative Chern-Simons extensions, Phys. Lett. B \textbf{451}, (1999) 73-76.
\bibitem{14} D.~S.~Kaparulin, I.~Yu.~Karataeva, S.~L.~Lyakhovich, Higher derivative extensions of 3d Chern-Simons models: conservation laws and stability, Eur. Phys. J. C \textbf{75}, (2015) 552.
\bibitem{15} V.~A.~ Abakumova, D.~S.~Kaparulin, S.~L.~Lyakhovich, Stable interactions in the higher derivative field theories of derived type, Phys. Rev.  D \textbf{99}, (2019) 045020. 
\bibitem{16} T.~Ehrgardt, K.~Rost, Resultant matrices and inversion of Bezoutians, Linear algebra and its applications \textbf{439}, (2013) 621-639.

\end{thebibliography}
\end{document}